# Quantum Encoding of Three-Dimensional Ligand Poses for Exhaustive Configuration Enumeration

Pei-Kun Yang

E-mail: peikun@isu.edu.tw; peikun6416@gmail.com

**Abstract**

Classical molecular docking is fundamentally constrained by the combinatorial growth of ligand translational and rotational degrees of freedom, rendering exhaustive pose enumeration infeasible on classical hardware. This work introduces a quantum-native formulation that encodes ligand occupancy on discretized three-dimensional grids and coherently generates the full ensemble of spatial configurations within a single quantum state. Multi-step translations and rotational transformations are controlled by ancillary qubits, enabling all symmetry-related configurations to be activated simultaneously. This framework provides a scalable foundation for quantum-accelerated virtual screening and is amenable to integration with quantum scoring approaches as quantum hardware continues to advance.

**Introduction**

Structure-based drug discovery depends on accurately estimating the interaction between a protein and its potential ligands [1-4]. When the ligand's binding pose is unknown, its three-dimensional translational position and rotational orientation must, in principle, be explored to identify energetically favorable configurations [5-7]. Electrostatic and van der Waals interactions are highly sensitive to interatomic separations, and even minor geometric deviations can introduce substantial errors in the estimated binding energy [8]. To maintain sufficient accuracy, ligand translations and rotations must therefore be sampled at fine spatial and angular resolutions. As the resolution increases, however, the number of possible ligand configurations grows combinatorially, rendering exhaustive evaluation computationally prohibitive on classical hardware.

Such calculations require extremely large-scale parallelism. Although modern GPUs provide substantial parallel throughput, the degree of parallelism required for exhaustive pose sampling remains astronomical. Quantum computers, through superposition and entanglement, offer the possibility of exploring vastly larger configuration spaces in parallel than classical devices [9-11]. This capability suggests a potential route for overcoming key computational bottlenecks in CPU- and GPU-based docking pipelines.

A central challenge nevertheless remains because force-field–based binding energy calculations depend explicitly on interatomic distances, which quantum hardware cannot efficiently compute natively. This limitation motivates the use of discretized grid-based energy evaluation schemes that avoid direct distance calculations [12]. Classical docking frameworks such as AutoDock address this challenge by precomputing electrostatic and van der Waals potentials on a discretized grid enclosing the receptor binding pocket. The grid defines the spatial domain over which protein–ligand interactions are evaluated, as illustrated in Figure 1 for the 1A30 protein–ligand complex [13]. Binding energies are then obtained by combining precomputed grid potentials with ligand grid charges and van der Waals atom-type occupancies. This approach converts continuous interaction calculations into discrete grid lookups, enabling efficient evaluation of ligand configurations.

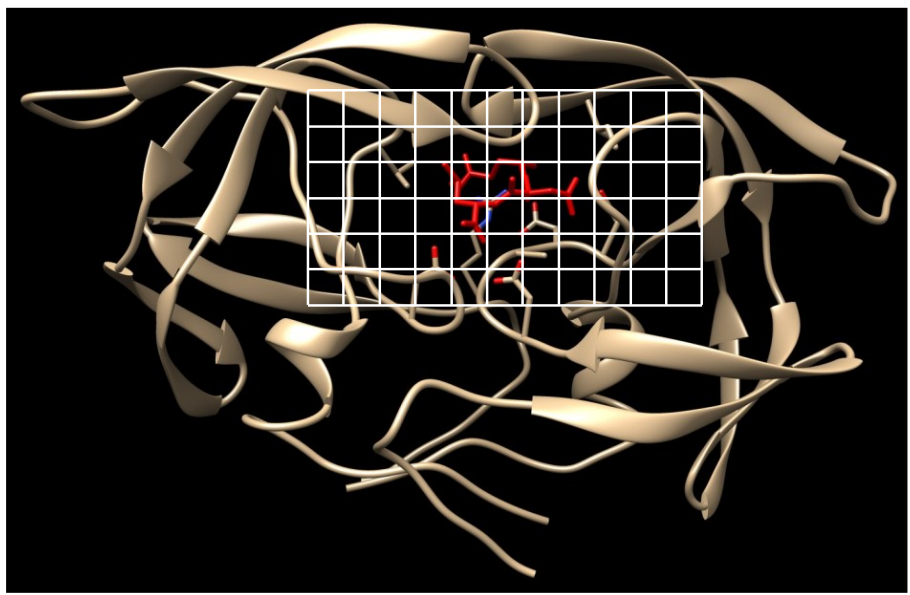

Figure 1. Structural representation of the 1A30 protein–ligand complex. The three-dimensional structure of the protein corresponding to PDB ID 1A30 is shown in cartoon representation, with the bound inhibitor highlighted in red and located in the central binding pocket. The overlaid grid illustrates a discretized spatial search region, representative of the grid-based evaluation domains commonly employed in classical docking frameworks such as AutoDock for estimating protein–ligand interactions.

Despite this acceleration, sampling precision remains a critical bottleneck. Even for moderate grid sizes, the combinatorial growth associated with pose sampling becomes severe. When each of the six geometric degrees of freedom, specifically the three translational degrees and the three rotational degrees, is discretized into one hundred possible values, the total number of ligand configurations reaches $10^{12}$. This trillion-scale search space lies far beyond the practical limits of classical exhaustive sampling and becomes increasingly prohibitive as finer spatial or angular resolutions are required. As a result, classical virtual screening pipelines face fundamental scalability constraints.

Quantum computing provides a fundamentally different approach to addressing this combinatorial explosion. The ability of quantum states to represent many configurations simultaneously has motivated growing interest in quantum-assisted drug discovery. Previous studies have explored quantum algorithms for molecular energy estimation and docking score prediction [14-16]. However, encoding the full set of ligand translations and rotations into quantum states typically requires substantial quantum resources, particularly when each pose must be constructed individually. The problem of generating all three-dimensional ligand translations and rotations directly and coherently within a quantum-native representation has not yet been systematically addressed.

In this work, we develop a quantum circuit architecture that performs complete three-dimensional ligand translations and rotations directly on grid-encoded ligand occupancy. Classical preprocessing is restricted to computing ligand grid charges and van der Waals atom-type occupancies, while all geometric manipulations are implemented using reversible quantum operations. Translational motion is realized through quantum shift operators controlled by ancillary registers, and rotational transformations are implemented using axis-specific quantum rotation units. Together, these components enable the full translational and rotational configuration space to be generated coherently within a single quantum state. The number of ligand poses represented grows exponentially with the number of ancillary qubits, allowing quantum hardware to explore configuration spaces far beyond the reach of classical sequential sampling.

By integrating grid-based molecular representations with quantum geometric operators, this work establishes a foundation for scalable quantum-assisted virtual screening. The proposed framework illustrates how quantum superposition can address one of the most computationally demanding components of docking workflows. As quantum hardware continues to advance, approaches of this kind may substantially expand the practical search space accessible in structure-based drug discovery.

**Theory**

**Quantum Ligand Translation.** Figure 2 presents a quantum framework for encoding ligand occupancy on a discretized spatial grid and implementing translational motion using quantum shift operators. The ligand's three-dimensional structure is first mapped onto a discrete lattice representation (Figure 2a), where each grid cell stores an occupancy or charge value. Although illustrated using a simplified 4 × 4 grid, the same encoding scheme extends directly to higher-resolution lattices employed in practical applications. Translational motion is realized by rigidly shifting the encoded occupancy pattern, with single-unit displacements along the $x$- and $y$-axes, as well as their combined action, shown in the corresponding panels.

Within this representation, each lattice coordinate is encoded as an ordered register of qubits, and translation is implemented through reversible binary increments of this register. A one-site displacement corresponds to incrementing the least significant qubit with carry propagation when required, while larger displacements, such as shifts of $2^p$ sites, initiate the increment at the corresponding bit position. These operations are implemented by the translation operators $U(2^0)$ and $U(2^p)$, as illustrated in Figures 2b and 2c.

The translation operators act as modular building blocks that combine additively. When the activation of each U($2^p$) is controlled by an ancillary register (Figure 2d), the resulting displacement equals the sum of all activated increments. Initializing the ancillary qubits with Hadamard gates prepares the control register in a uniform superposition over all binary displacement patterns, enabling the circuit to coherently generate all possible translation distances simultaneously within a single quantum state. This parallel generation eliminates the need for classical enumeration of individual shifts.

This construction generalizes naturally to three spatial dimensions by encoding the *z*-, *y*-, and *x*-coordinates into independent quantum shift registers $\boldsymbol{T}^z$, $\boldsymbol{T}^y$, and $\boldsymbol{T}^x$, with separate ancillary registers specifying translation distances along each axis (Figure 2e). For the *z*-direction, an ancillary register of $mz$ qubits determines whether the operators U($2^0$), U($2^1$), …, U($2^{mz-1}$) are applied, with analogous structures for the *y*- and *x*-axes. When shift operators along all three axes act on their respective registers, the encoded ligand spans every displacement from 0 to $2^{mz-1}$ along *z*, 0 to $2^{my-1}$ along *y*, and 0 to $2^{mx-1}$ along *x*. Initializing all ancillary registers in uniform superposition produces the full set of $2^{mz+my+mx}$ translated ligand configurations coherently, providing a scalable extension of the one-dimensional quantum shift operator and enabling quantum-parallel pose enumeration beyond the reach of classical simulation.

(a)

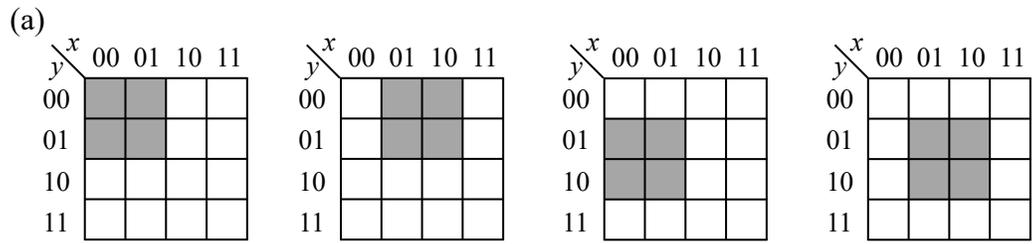

(b)

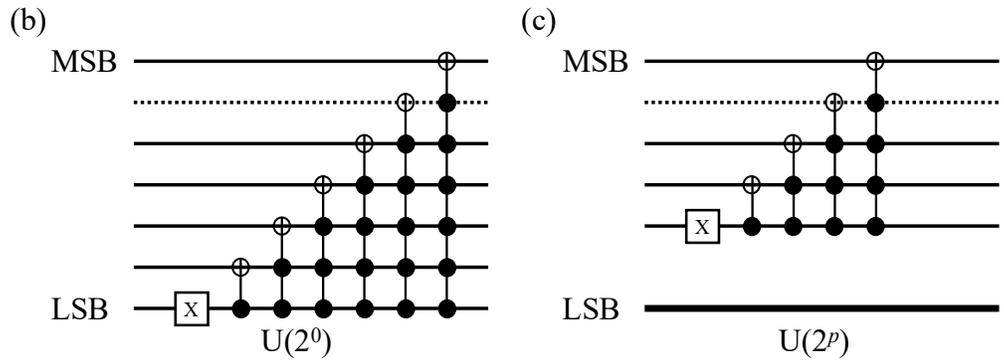

(c)

(d)

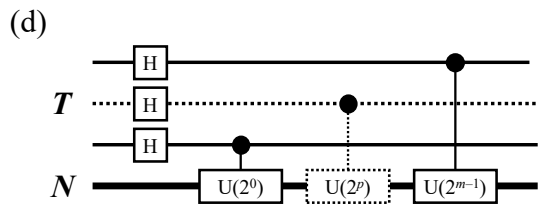

(e)

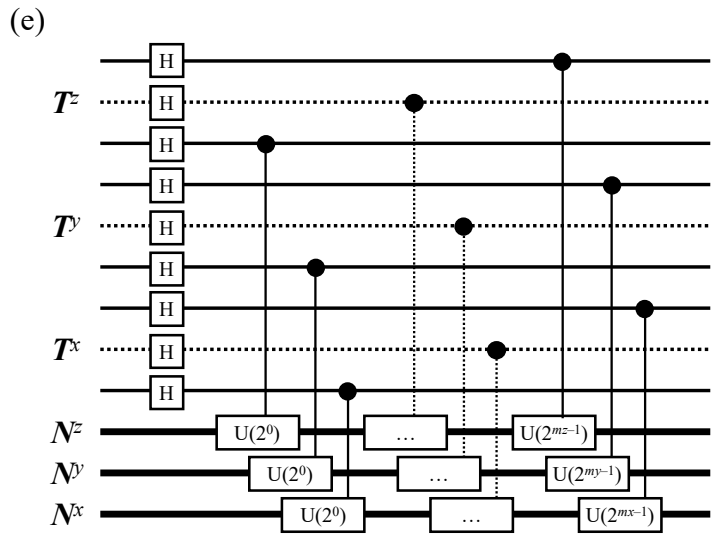

Figure 2. Ligand translation on a discretized grid and its quantum implementation. (a) The original ligand configuration together with its translations by one grid unit along the *x*-axis, the *y*-axis, and both axes. (b)–(c) Quantum shift operators that implement translations of the ligand position register by $2^0$ and $2^p$ lattice sites, respectively. (d) An extended construction in which an *m*-qubit ancillary register *T*, initialized by Hadamard gates, coherently controls a set of quantum shift operators acting on the ligand position register *N*, enabling superpositions of all translation distances from 0 to $2^m - 1$. (e) A three-dimensional translation circuit defined on a discretized spatial grid, where the *z*-, *y*-, and *x*-coordinates are encoded by independent quantum shift registers $T^z$, $T^y$, and $T^x$, and controlled shift operators apply the corresponding translations along each axis.

**Coordinate-Swap Transformations on Discretized Ligand Grids.** The geometric effect of exchanging coordinate indices is illustrated in Figure 3a, where a two-dimensional discretized ligand grid is labeled by four distinct grid elements. Interchanging the *x* and *y* coordinates permutes the positions of these elements, corresponding to a reflection across the diagonal of the grid and demonstrating how coordinate permutations reshape the spatial arrangement of grid occupancy.

The quantum implementation of this transformation is shown in Figure 3b, where the *x* and *y* indices are encoded in two equally sized qubit registers. The coordinate-swap operation is realized through a sequence of pairwise SWAP gates acting between corresponding qubits in the two registers. To enable coherent access to multiple spatial configurations, the construction in Figure 3c introduces an ancillary control qubit that determines whether the swap is applied. When the ancilla is in the $|0\rangle$ state, the configuration remains unchanged; when it is in the $|1\rangle$ state, the swap operation is activated. Initializing the ancilla in superposition generates a coherent quantum state containing both the swapped and unswapped ligand configurations.

The framework extends naturally to three spatial dimensions, as shown in Figure 3d, where coordinate swaps are applied to the $(x, y)$, $(y, z)$, and $(z, x)$ axis pairs. Each pairwise exchange is independently controlled, allowing the circuit to generate all eight possible permutations of the coordinate triplet $(x, y, z)$. When the corresponding control qubits are prepared in superposition, all eight volumetric configurations are represented simultaneously within a single coherent quantum state. This construction systematically enumerates the coordinate-exchange symmetries relevant to three-dimensional ligand orientation.

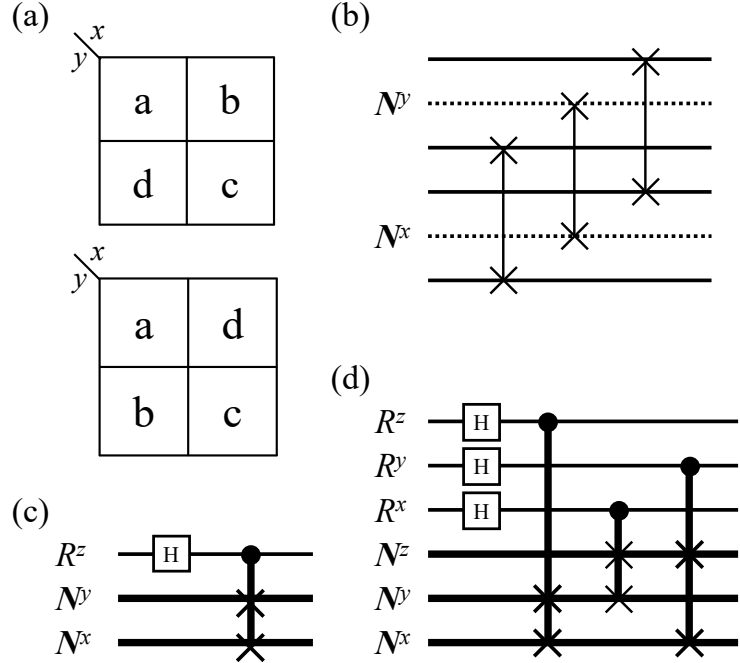

Figure 3. Coordinate-swap transformations on discretized ligand grids and their quantum implementations. (a) A two-dimensional discretized grid labeled by four distinct elements, illustrating the effect of exchanging the $x$ and $y$ coordinates, which permutes the positions of the grid elements and reflects the configuration across the diagonal. (b) The quantum circuit that implements this $x$–$y$ coordinate swap by exchanging the corresponding quantum registers. (c) An extended construction in which an ancillary control qubit, initialized by a Hadamard gate, coherently selects between the swapped and unswapped configurations, generating their superposition within a single quantum state. (d) The three-dimensional extension, where independent control qubits govern coordinate swaps between the $(x, y)$, $(y, z)$, and $(z, x)$ register pairs. Each control qubit determines whether the corresponding axis pair is swapped or left unchanged, yielding all eight symmetry-related volumetric configurations coherently within a single quantum circuit.

**Quantum Realization of Discrete Grid Rotations.** In the illustrative example shown in Figure 4a, a 4 × 4 discretized grid is partitioned into four labeled regions, and a 90° rotation about the $z$-axis cyclically permutes these regions. The same transformation is implemented quantum mechanically by rearranging the binary-encoded coordinate registers according to this permutation, as depicted in Figure 4b. A controlled version of the rotation is obtained by coupling the permutation operation to an ancillary qubit, as illustrated in Figure 4c. When this control qubit determines whether the rotation is applied, preparing it in superposition yields a coherent quantum state that contains both the original and rotated ligand configurations simultaneously, enabling multiple orientations to be accessed within a single circuit execution.

The framework extends naturally to three spatial dimensions, as shown in Figure 4d, where independent 90° rotations about the *z*-, *y*-, and *x*-axes are governed by separate control qubits. Each control determines whether the corresponding axial rotation is applied or omitted. When all control qubits are prepared in superposition, the circuit coherently generates all eight combinations of axis-aligned rotations, providing a compact and scalable mechanism for producing the complete discrete rotational ensemble.

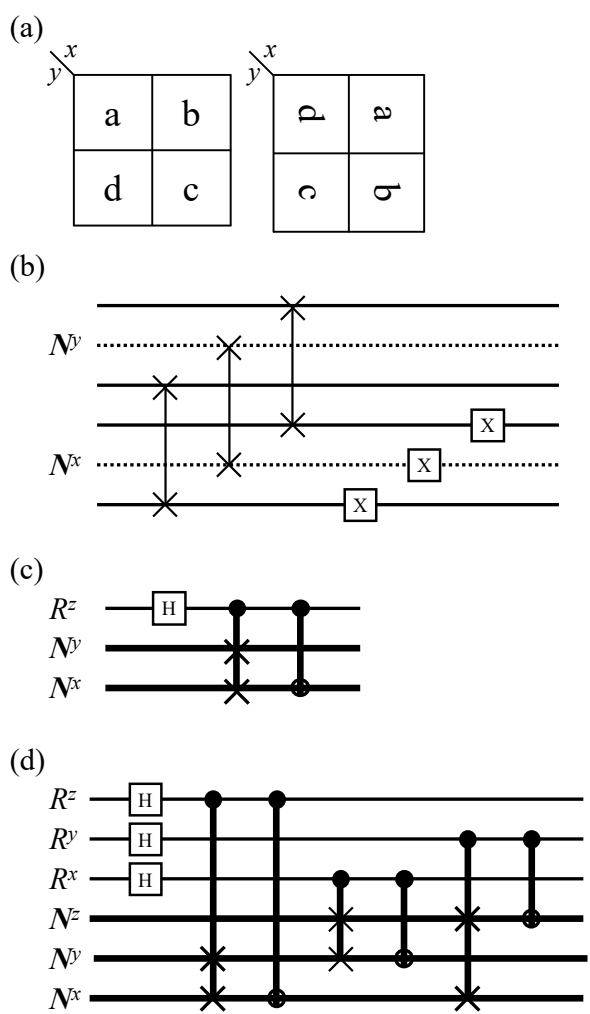

Figure 4. Discrete grid rotations and their quantum implementations. (a) A discretized two-dimensional grid labeled by distinct elements, together with the configuration obtained

after a 90° rotation about the *z*-axis, illustrating how a discrete axial rotation permutes the grid positions. (b) The quantum circuit that implements this rotation by permuting the coordinate-register qubits and applying the corresponding bit-flip operations. (c) An extended construction in which an ancillary control qubit, initialized by a Hadamard gate, coherently selects between the rotated and unrotated configurations, generating their superposition within a single quantum state. (d) The three-dimensional generalization, where independent control qubits govern discrete 90° rotations about the *z*-, *y*-, and *x*-axes. Each control qubit determines whether the corresponding axial rotation is applied or omitted, yielding all eight symmetry-related rotational configurations coherently within a single quantum circuit.

**Unified Quantum Circuit for Three-Dimensional Translation, Swapping, and Rotation.** The quantum representation of ligand motion is formulated by encoding its spatial occupancy on a discretized three-dimensional grid, with the *z*-, *y*-, and *x*-coordinates stored in independent quantum position registers. To enumerate all possible displacements, the ancillary translation registers $T^z$, $T^y$, and $T^x$ are initialized in uniform superposition, while controlled operators $U(2^p)$ apply increments of magnitude $2^p$ along the corresponding spatial axis. This construction generates a coherent superposition containing all $2^{mz+my+mx}$ translated ligand configurations.

Discrete geometric symmetries are incorporated through a set of axis-specific control registers $R^z$, $R^y$, and $R^x$, which govern symmetry operations associated with the *z*-, *y*-, and *x*-axes, respectively. When activated, each register controls a discrete 90° rotation about its corresponding axis, implemented as a cyclic permutation of the encoded grid coordinates. The same control registers also govern coordinate-exchange operations, such that activation of $R^z$, $R^y$, and $R^x$ conditionally applies controlled-swap operations that interchange the (*x*, *y*), (*y*, *z*), and (*z*, *x*) coordinate registers, respectively. Each axis-pair exchange can be independently activated or omitted, allowing all eight possible combinations of axis-aligned rotations and coordinate permutations to be generated coherently. Preparing the control registers in superposition enables the circuit to represent the identity operation together with all discrete rotational and coordinate-exchange configurations simultaneously within a single coherent quantum state.

By integrating translation, coordinate exchange, and rotation into a unified quantum architecture, the circuit spans the complete family of symmetry-related ligand configurations in three dimensions. When all ancillary registers are prepared in superposition, the circuit produces the entire ensemble of translated, swapped, and rotated ligand states within a single execution. This unified construction provides a scalable foundation for downstream quantum procedures, such as receptor–ligand energy evaluation.

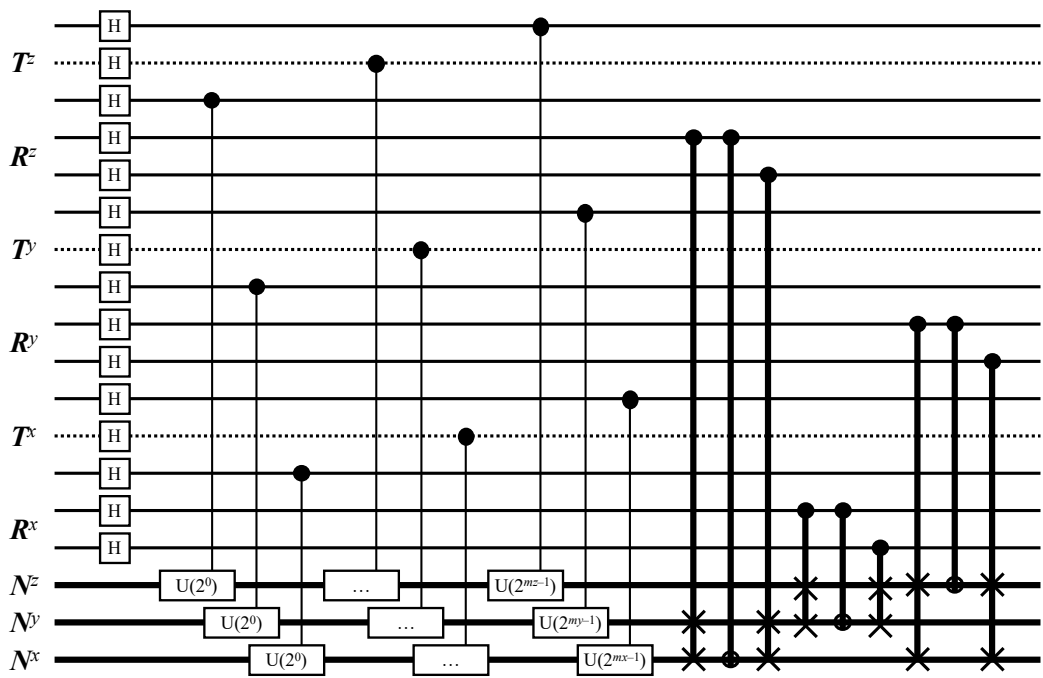

**Figure 5. Unified quantum circuit for exhaustive enumeration of ligand translations, coordinate exchanges, and axis rotations.** The translation registers $T^z$, $T^y$, and $T^x$ are initialized into uniform superposition by Hadamard gates, where the $mz$, $my$, and $mx$ qubits encode all discrete displacements along the three spatial directions. Independent rotation-control registers $R^z$, $R^y$, and $R^x$ govern the conditional application of coordinate-exchange operations and discrete 90° axis rotations about the corresponding axes. Each control register is prepared in superposition, enabling the associated symmetry operation to be either applied or omitted coherently. When all translation and rotation-control registers are initialized in superposition, the circuit generates every combination of displacement, coordinate exchange, and axis-aligned rotation within a single coherent quantum state, spanning the complete symmetry-related configuration space of the ligand.

## Conclusion

This work introduces a unified quantum framework for implementing translational and rotational transformations of ligand configurations on discretized spatial grids. By encoding ligand occupancy into qubit-based spatial registers, the proposed circuits coherently generate the full set of ligand translations and rotations within a single quantum state. This formulation provides a scalable, intrinsically parallel approach to the geometric exploration required for structure-based virtual screening. The results presented here establish a foundational building block for quantum-accelerated virtual screening. As quantum hardware continues to advance, the circuits developed in this work can be integrated with quantum scoring and optimization procedures, opening a path toward fully quantum-native workflows capable of exploring spatial degrees of freedom far beyond the practical limits of classical methods.